\documentstyle[preprint,aps]{revtex}
\begin{document}
\draft
\newfont{\form}{cmss10}
\newcommand{\e}{\varepsilon}
\renewcommand{\b}{\beta}
\newcommand{\unity}{1\kern-.65mm \mbox{\form l}}%
\newcommand{\D}{D \raise0.5mm\hbox{\kern-2.0mm /}}
\newcommand{\A}{A \raise0.5mm\hbox{\kern-1.8mm /}}
\def\pmb#1{\leavevmode\setbox0=\hbox{$#1$}\kern-.025em\copy0\kern-\wd0
\kern-.05em\copy0\kern-\wd0\kern-.025em\raise.0433em\box0}

\def\D{\hbox{\hbox{${D}$}}\kern-1.9mm{\hbox{${/}$}}}
\def\kbar{\hbox{$k$}\kern-0.2truecm\hbox{$/$}}
\def\nbar{\hbox{$n$}\kern-0.23truecm\hbox{$/$}}
\def\pbar{\hbox{$p$}\kern-0.18truecm\hbox{$/$}}
\def\nhbar{\hbox{$\hat n$}\kern-0.23truecm\hbox{$/$}}
\newcommand{\dif}{\hspace{-1mm}{\rm d}}
\newcommand{\dil}[1]{{\rm Li}_2\left(#1\right)}
\newcommand{\diff}{{\rm d}}

\title{Discontinuous behaviour of perturbative Yang-Mills theories
in the limit of dimensions $D\to 2$}
\author{A. Bassetto}
\address{Dipartimento di Fisica ``G.Galilei", Universit\`a di Padova,
Via Marzolo 8, 35131
Padova, Italy\\
INFN, Sezione di Padova, Italy}
\author{R. Begliuomini and G. Nardelli}
\address{Dipartimento di Fisica, Universit\`a di Trento,
38050 Povo (Trento), Italy \\ INFN, Gruppo Collegato di Trento, Italy}

\maketitle
\begin{abstract}
We calculate in dimensions $D=2+\epsilon$ and in light-cone gauge (LCG)
the perturbative ${\cal O}(g^4)$
contribution to a rectangular Wilson loop in the $(t,x)$-plane coming
from diagrams with a self-energy correction in the vector propagator.
In the limit $\epsilon \to 0$ the result is finite, in spite of
the vanishing of the triple vector vertex in LCG, and provides
the expected agreement with the analogous calculation in Feynman
gauge.
\end{abstract}

\noindent Padova preprint DFPD 99/TH/13, April 1999.
\vfill\eject

\narrowtext

\section{introduction}

$SU(N)$ Yang-Mills (YM) theories exhibit peculiar and
interesting features in 1+1 dimensions ($D=2$). The reduction
from four to lower dimensions  entails indeed tremendous simplifications,
so that many problems can be faced, and often exactly solved
\cite{wit},\cite{doug},\cite{basgri}. For instance exact evaluations
of vacuum to vacuum
amplitudes of Wilson loop operators, that, for a suitable choice of contour
and in a particular limit, provide the potential between a static ${\rm q}
{\rm \bar q}$ pair \cite{poly72},\cite{fish},\cite{wils74}, can be obtained.

YM theories without fermions in 1+1 dimensions
are considered free theories, apart from topological
effects. This feature looks apparent when choosing an axial gauge.
However, either when matter fields are introduced, or in Wilson Loop
calculations, the perturbative  $1+1$
dimensional theory exhibits dramatic infrared (IR) singularities which need
to be regularized. Unfortunately the results appear to be dependent on such
regularization procedures, even when they concern gauge invariant
quantities \cite{tellu}.

In light cone gauge (LCG) the (IR) singular behaviour is particularly
apparent in the vector propagator, where the gauge pole conspires
with the usual Feynman singularity to produce a double pole \cite{thooft}.

A Cauchy principal value (CPV) prescription for this IR singularity
has often been advocated \cite{call76}.
It emerges quite naturally if the theory is quantized on the light cone
surface $x^+=0$ \cite{Bas4}.

On the other hand such a recipe is at odds with Wick's rotation.
In ref.\cite{Wu} a causal prescription for the double pole has been proposed,
which is nothing but the one suggested
years later by Mandelstam and Leibbrandt (ML) \cite{ML},
when restricted to $1+1$ dimensions. This prescription
follows from equal-time quantization \cite{Bas5}
and is mandatory in order to renormalize the theory in 1+3
dimensions\cite{Bas3},\cite{Bas4}.

In view of the above-mentioned results and of the
fact that ``pure'' YM theories do not immediately look free in Feynman gauge,
a systematic investigation has been undertaken to clarify their properties
when the two dimensional picture is reached starting
from higher dimensions.

Since no exact solutions are available beyond $D=2$, the investigation
has been focussed on perturbative calculations, looking for consistency
checks, in particular testing the gauge invariance of the theory
which holds order by order in the coupling constant expansion.

Recalling that perturbative S-matrix elements cannot be
consistently defined in non-Abelian gauge theories, owing
to their (IR) singular mass-shell behaviour,
the natural gauge invariant quantities to be
considered are Wilson loops.

A first test of gauge invariance in $1+3$ dimensions has been performed
in refs.\cite{Korc,Bas2} by calculating at ${\cal
O}(g^4)$, both in Feynman and in light-cone gauge with ML prescription,
a rectangular Wilson loop with
light-like sides, directed along the vectors $n_\mu = (T, - T)$,
$n^*_\mu = (L, L)$ and parametrized according to the equations:

\begin{eqnarray}
\label{uno}
C_1 &:& x^\mu (t) = n^{* \mu} t,\nonumber\\
C_2 &:& x^\mu (t) = n^{* \mu} + n^\mu t,\nonumber\\
C_3 &:& x^\mu (t) = n^\mu + n^{* \mu}( 1-t), \nonumber\\
C_4 &:& x^\mu (t) = n^\mu (1 - t), \qquad 0 \leq t \leq 1.
\end{eqnarray}

In order to perform the test, dimensional regularization ($D=2\omega$)
was used for both UV and IR singularities. Full consistency between Feynman and
light-cone gauge with the ML prescription was obtained.

Since results in $2\omega$ dimensions were available, in view of the peculiar
features of Yang-Mills theories in 2 dimensions mentioned above, the interest
arose in knowing the outcome of the check in the limit $\omega \to 1$.
The following unexpected results were obtained in \cite{Bas1}.

The ${\cal O}(g^4)$ perturbative loop expression in $d= 1+(D - 1)$
dimensions is {\it finite} in the limit $D\to 2$. The loop
expression is a function
only of the area $n\cdot n^{*}$ for any dimension $D$ and exhibits
also a dependence
on $C_A$, the Casimir constant of the adjoint representation.

In LCG this dependence
comes from two sources:
\begin{itemize}
\item  diagrams with two crossed propagators
(colour factor $C_F(C_F - C_A/2)$,
$C_F$ being the Casimir constant of the fundamental representation);
\item a genuine contribution to the Wilson loop  proportional to $C_F C_A$
coming from the one-loop correction to the vector propagator (self-energy
diagram).
\end{itemize}

We shall concentrate our interest on the contribution due to this self-energy
diagram. At a first sight, it is surprising, since, in  1+1 dimensions,
there is no triple vector vertex in axial gauges. What happens is that
the vanishing strength of the vertex at $D=2$ matches
the self-energy loop singularity,
eventually producing  a finite result. Feynman diagrams with a triple vertex
but no loops tend instead smoothly to zero when inserted in the Wilson contour.

We notice that no ambiguity affects the ${\cal O}(g^4)$
gauge invariant result,
which is finite; in addition the presence of $C_A$
cannot be re-absorbed by a redefinition of the coupling,
that, while unjustified on general grounds, would also turn out
to be dependent on the area of the loop.

In order to clarify whether the appearance of $C_A$ in the
maximally non-Abelian term is indeed a pathology, one should
examine the potential $V(2L)$ between a
``static" ${\rm q} {\rm\bar q}$ pair in the fundamental representation,
separated by a distance $2L$. Therefore in ref.\cite{bello}
we have considered a different
Wilson loop, {\it viz} a rectangular loop with
one side along the space direction and one side along the time direction,
of length $2L$ and $2T$ respectively. Eventually the limit $T \to \infty$
at fixed $L$ is to be taken: the potential $V(2L)$  between the
quark and the antiquark is indeed
related to the value of the corresponding Wilson loop
amplitude ${\cal W}(L,T)$ through the equation \cite{ALTE}
\begin{equation}
\lim_{T\to\infty}{\cal W}(L,T)=const.\  e^{-2i T V(2L)}\ .
\label{potential}
\end{equation}

The crucial point to notice in eq.(\ref{potential}) is that
dependence on the Casimir
constant $C_A$ should cancel at the leading order when $T\to \infty$ in any
coefficient of a perturbative expansion of the potential with
respect to coupling constant. This criterion has often been used
as a check of gauge invariance \cite{Bas4}.

In ref.\cite{bello} the calculation has been performed in Feynman
gauge, obtaining the following results.

For $D>2$ the ${\cal O}(g^4)$ perturbative expression of the loop
depends, besides on the area,
also on the ratio $\beta=L/T$. As we are eventually interested
in the large-$T$ behavior, we have always considered the region $\beta<1$;
moreover we have chosen $D=2+\epsilon $ with
a small $\epsilon >0.$

As long as $D>2$, agreement with Abelian-like
time exponentiation (ALTE) occurs
in the limit $T\to \infty$, with a pure $C_F$-dependence in the
leading coefficient. Consistency of all previous
results \cite{Bas4} in higher dimensions is thus re-established.

The limit $D\to 2$ for
$\beta=0$
{\it exactly} reproduces the gauge invariant result obtained in
ref.\cite{Bas1} for a loop of the same area with light-like sides;
thereby we enforce the argument that in two dimensions a pure
area behaviour is expected, no matter the orientation and the
shape of the loop. What may be surprising is that the term,
which in LCG corresponds to the self-energy
correction, exhibits, in the limit, a pure area dependence on its own.

However, in two dimensions at
${\cal O}(g^4)$, a $C_A$-dependence is
definetely there and agreement with ALTE
is lost. Actually this behaviour at $D=2$ persists at any order of $g$ and
affects the sum of the perturbative series \cite{stau},\cite{basgri}.

A peculiar feature of the light-cone gauge in $2$ dimensions is that individual
Wilson loop diagrams do not exhibit any singularity; hence there is  
no need of dimensional regularization.

In ref. \cite{Bas7},   a
${\cal O}(g^4)$ perturbative calculation of the Wilson loop in LCG with ML
prescription, for a rectangular loop with sides $2T\times 2L$ lying in the
$x^0\times x^1$ axes, was performed at $D=2$.
No agreement occurs with the result one finds in ref.\cite{bello} when taking
the limit $D\to 2$. The source of such a discrepancy is rooted in the mentioned
self-energy diagram contribution, which is obviously missing at $D=2$, but
provides a finite term in the limit
$D\to 2$, thereby producing a discontinuity in the theory \cite{Bas1}.

The purpose of this paper is  to check explicitly this property by evaluating
in LCG the relevant discontinuity for the Wilson loop of ref. \cite{Bas7}.
 We confirm  that the missing term comes from the
diagram with a self-energy corrected propagator, evaluated at $D=2+\epsilon$,
when eventually taking the limit $\epsilon \to 0$. We thereby reproduce
for a space-time contour the phenomenon in LCG found in
ref.\cite{Bas1} for a contour with light-like sides.
Actually, from
the computation of the self-energy diagram at $D>2$, we find, as an extra
bonus, that its contribution vanishes for $\epsilon >0$ in the limit
$T\to \infty$ with the same ``universal'' factor $T^{4-4\omega}$ we
have obtained in ref.\cite{bello} for the maximally non-Abelian
contributions \cite{tay}.

The limits $T \to \infty$ and $\epsilon \to 0$  {\it do not commute}.

\section{The calculation}

We recall some basic notions and notations.
We consider, as in ref.\cite{bello}, the closed path $\gamma$
parametrized by the following  four segments
$\gamma_i$,
\begin{eqnarray}
\gamma_1 &:& \gamma_1^\mu (s) = (sT, L)\ ,\nonumber\\
\gamma_2 &:& \gamma_2^\mu (s) = (T,-sL)\ ,\nonumber\\
\gamma_3 &:& \gamma_3^\mu (s) = (-sT, -L)\ , \nonumber\\
\gamma_4 &:& \gamma_4^\mu (s) = (-T, sL)\ , \ \ \qquad -1 \leq s \leq 1.
\label{path}
\end{eqnarray}
describing a  (counterclockwise-oriented) rectangle
centered at the origin of the plane ($x^1,x^0$),
with length sides $(2L,2T)$, respectively.

The perturbative expansion of the Wilson loop  is
\begin{equation}
{\cal W}_\gamma (L,T) = 1 + {1\over N}\sum_{n=2}^\infty (ig)^n \oint_\gamma
dx_1^{\mu_1} \cdots \oint_\gamma dx_n^{\mu_n}\theta( x_1 >\cdots >x_n )
{\rm Tr} [ G_{\mu_1 \cdots \mu_n} (x_1,\cdots ,x_n)]\ ,
\label{wilpert}
\end{equation}
where $ G_{\mu_1 \cdots \mu_n} (x_1,\cdots ,x_n)$ is the Lie algebra valued
$n$-point Green function, and   the Heavyside $\theta$-functions order the
points
$x_1,\cdots ,x_n$ along the integration path $\gamma$.

It is easy to show that the perturbative expansion of ${\cal W}_\gamma$ is an
even power series in the coupling constant, so that we can write
\begin{equation}
{\cal W}_\gamma (L,T)= 1+g^2 {\cal W}_2 + g^4 {\cal W}_4 + {\cal O}(g^6)\ .
\label{pert}
\end{equation}

To have a sensitive check of gauge invariance, one has to consider at least the
order $g^4$, ({\it i.e.} one has to evaluate ${\cal W}_4$), as this is the
lowest order where genuinely non-Abelian $C_FC_A$ contributions may appear.
In turn, in the calculation of ${\cal W}_4$, only the so called maximally
non-Abelian contribution  ${\cal W}_4^{na}$ need to be evaluated, that in our
case comes from the terms proportional to $C_FC_A$. The Abelian contribution,
proportional to $C_F^2$,  can be easily obtained thanks to the Abelian
exponentiation theorem \cite{tay}.

The diagrams contributing to ${\cal W}_4^{na}$ can be grouped into three
families: a) crossed diagrams (${\cal C}_{(ij)(kl)}$), with a double gluon
exchange in which the two (crossed)  propagators join the sides $(ij)$ and
$(kl)$ of the contour $\gamma$; b) spider diagrams (${\cal S}_{ijk}$), which are
obtained by attaching a three point Green function at the tree level to the
sides $(ijk)$ of the loop; c) bubble diagrams (${\cal B}_{ij}$) , that are
single
exchange diagrams in which the gluon propagator, corrected by a self-energy
term,  joins the sides $(ij)$ of the contour.

In arbitrary dimensions,  the calculation of the  Wilson loop
is much more awkward in LCG than in covariant gauge, due to
a more complicated form of the vector propagator. However, when considering the
$D\to 2$ limit, diagrams in LCG have much better analyticity properties
in $\omega$ than the ones in Feynman gauge.
The vector propagator in LCG with
ML prescription is a tempered distribution at $D=2$,  at odds with the one
in Feynman gauge. Moreover it is summable along
the (compact) loop contour.

Due to this property, we can conclude that all the maximally non-Abelian
contributions arising from diagrams with crossed propagators
sum to an expression that, in the
limit $D\to 2$, reproduces the result of ref. \cite{Bas7}, namely
\begin{equation}
\label{croci}
{\cal W}^{cr}=C_AC_F{(LT)^2\over 3}\ \ .
\end{equation}

\smallskip

Now we consider the contribution ${\cal W}^{bub}$
coming from bubble diagrams.  In LCG and on the plane $x^0\times x^1$,  the only
non-vanishing component of the two point Green function $\Delta_{\mu\nu}$ at the
order ${\cal O}(g^2)$ is $ \Delta_{++}(x)\equiv \Delta(x)$, that reads, at
$x_\perp=0$ \cite{Bas2},
\begin{equation}
\label{self}
\Delta(x)=-\frac{g^2}{8\pi^{2\omega}}C_A\frac{(x^-)^2}{(-x^2+i\e)^{2\omega-2}}
f(\omega)\ ,
\end{equation}
\begin{equation}
\label{fd}
f(\omega)=\frac{1}{(2-\omega)^3}\left[\frac{\Gamma^2(3-\omega)\Gamma(2\omega
-3)}{\Gamma(5-2\omega)}
 -{\Gamma(\omega-1) \Gamma(\omega) (10\omega^2 -19\omega + 10)\over
4(2\omega-3)(2\omega-1)}\right] \ .
\end{equation}
Following the notations of ref. \cite{bello}, there are 10 topologically
inequivalent bubble diagrams.
However, due to the symmetry of the Green function and to the symmetric choice
of the contour, only six of them  are independent, and the ${\cal O}(g^4)$
contribution  to the Wilson loop arising from bubble diagrams can be
written as
\begin{equation}
\label{bub1}
{\cal W}^{bub}= 2({\cal B}_{11} + {\cal B}_{22} +{\cal B}_{13}+ {\cal B}_{24}
+ 2{\cal B}_{12}+ 2{\cal B}_{14}) \ ,
\end{equation}
where each single contribution ${\cal B}_{ij}$ can be calculated by replacing
eqs. (\ref{path}), (\ref{self}) in the formula
\begin{equation}
\label{bubbola}
{\cal B}_{ij}=-{1\over 2}g^2 C_F \int_{-1}^1 ds \int_{-1}^1 dt \Delta_{\mu\nu}
(\gamma_i (s)- \gamma_j(t)) \dot \gamma^\mu_i (s)\dot \gamma^\nu_j (t)\ ,
\end{equation}
where the dot denotes derivative with respect to the variable parametrizing the
segment.

The calculation being standard, we shall report only the final result
\begin{eqnarray}
\label{bub2}
&&{\cal W}^{bub}= {C_FC_A\over \pi^{2\omega} }f(\omega) (LT)^2
(2L)^{4-4\omega}\left\{e^{-2i\pi\omega }\beta^{4\omega-6}\left[ {1\over
(7-4\omega)(8-4\omega)}\right.\right.\nonumber\\
&&\times\biggl(1-(8-4\omega)
_2F_1(2\omega-2,2\omega-7/2;2\omega-5/2;\beta^2)+(7-4\omega)
(1-\beta^2)^{3-2\omega}\biggr)\nonumber\\
&& \left.
-{1\over(3-2\omega)(4-2\omega)}\left(1-(1-\beta^2)^{4-2\omega}\right)
+ {5-2\omega\over(6-4\omega)(4-2\omega)} \left(1-(1-\beta^2)^{3-2\omega}\right)
\right]\nonumber\\
&&+e^{-2i\pi\omega}
\beta^{4\omega-4}\left[{(1-\beta^2)^{3-2\omega} \over
(3-2\omega)(4-2\omega)}- { _2F_1(2\omega-2,
2\omega-5/2;2\omega-3/2;\beta^2)\over (5-4\omega)}\right.\nonumber\\
&&\left. -  _2F_1(2\omega-2,1/2;3/2;\beta^2){ { } \over  { } }\right]
+i\beta {\sqrt{\pi} (\omega-2)\Gamma(2\omega -7/2)\over
\Gamma(2\omega -2)}  \nonumber\\
&&\left.  -e^{-2i\pi\omega} {\beta^{4\omega-2}\over 3}
  {_2F_1}(2\omega-2,3/2;5/2;\beta^2) + { \beta^2\over (7-4\omega)}\right\}\ ,
 \end{eqnarray}
where $\beta = L/T$.

Some comments are here in order. First of all there is a dependence
on the dimensionless ratio $\beta$, besides the area, at variance with
the analogous result in LCG for
the rectangle of light-like sides. However, in the equation above,
one can easily check
that the quantity ${\cal W}^{bub}/(LT)^2$ is not singular for $\beta\to 0$.
Actually eq.(\ref{bub2}) exhibits, for $\omega >1$, the expected
damping factor $T^{4-4\omega}$ in the large-$T$ limit.

In the limit $\omega\to 1$  the dependence on $\beta$ disappears and the
pure area law is recovered: ${\cal W}^{bub} =C_FC_A (LT/\pi)^2$.
This is exactly the ``missing'' term to be added to the expression of ref.\cite
{Bas7} to obtain the final result for
the maximally non-Abelian contribution to the perturbative ${\cal O}(g^4)$
Wilson loop in the limit $D\to 2$,
\begin{equation}
\label{finale}
{\cal W}_4^{na} = C_FC_A \left({LT\over \pi}\right)^2 \left[ 1 + {\pi^2\over
3}\right]\ .
\end{equation}

Equation (\ref{finale}) is in full agreement not only with ref. \cite{bello},
where an anologous Wilson loop was calculated in Feynman gauge, but also with
ref. \cite{Bas1}, where the loop was oriented in a different direction.
Moreover, in LCG, different families of diagrams (``crossed'' and ``bubble''
diagrams) give the same contribution ($C_FC_A {(LT)^2 \over 3}$
and $C_FC_A \left({LT\over \pi}\right)^2$ respectively) no matter the
orientation of the loop:
remarkably, invariance under area-preserving diffeomorphisms is
recovered in the limit $D\to 2$, even when the Wilson loop
is first evaluated  in higher dimensions, and then the limit $D\to
2$ is taken.

In turn the result above implies that ``spider'' diagrams, namely diagrams with
a triple vector vertex, cannot contribute in the limit $D \to 2$.
This is not surprising, as the same phenomenon occurred in ref.
\cite{Bas1}, although for a different contour (contour with light-like sides).

In order to support this conclusion,
we show that the relevant
three point Green
function at ${\cal O}(g)$, vanishes when $D \to 2$.

To this aim, let us consider the three point Green function
${\cal V}_{\mu\nu\rho} (x,y,z)$. Due to the LCG choice, its only non-vanishing
component when considering the loop in the $x^0\times x^1$ plane is ${\cal
V}(x,y,z) = {\cal V}_{+++}(x,y,z)$; up to an irrelevant
multiplicative constant, it is given by
\begin{equation}
\label{gamma}
{\cal V}(x,y,z)=\int d^{2\omega}\zeta {\partial\over \partial z^\alpha}
\left[{\partial\over \partial x^\alpha}{\partial\over \partial y^+}-
{\partial\over \partial y^\alpha}{\partial\over \partial x^+}\right]
F(x-\zeta)F(y-\zeta) G(z-\zeta)
\end{equation}
$$ +{\rm cycl.\  perm.} \ \{x,y,z\}\ \equiv ({\cal V}_1-{\cal V}_2)
+{\rm cycl.\  perm.} \ \{x,y,z\}\ .$$

Here the index $\alpha$ runs over the transverse components and the functions
$G$ and $F$ are the following Fourier transforms
\begin{equation}
\label{g}
G(x)=\int d^{2\omega} p {e^{ipx}\over p^2 +i\e}=-\pi^\omega
\Gamma(\omega-1) \left(-{x^2\over 4} + i\e\right)^{1-\omega},
 \end{equation}
\begin{equation}
\label{f}
F(x)=\int d^{2\omega} p {e^{ipx}\over (p^2 +i\e) (p^++i\e p^-)}=
-i\pi^\omega \Gamma(\omega-1)\int_0^{x_+} d\rho \left({x_\perp^2 - 2 x_-
\rho\over 4} + i\e\right)^{1-\omega}.
 \end{equation}

Let us consider, for instance, the first term in eq. (\ref{gamma}), that
we call ${\cal V}_1$. Using
standard Feynman integrals techniques, integrations over momenta and over the
intermediate point $\zeta$ can be performed, so that ${\cal V}_1$ can be
rewritten, after some convenient change of variables, as
\begin{eqnarray}
\label{gamma1}
{\cal V}_1&=&{i\pi^\omega (4 \pi)^{3\omega}\over 8}\Gamma(2\omega-1)
(\omega-1)\!\! \int_0^1 \!\! d\xi d\eta d\mu \,  \eta [\mu(1-\mu)]^{\omega -1}
\int_0^\infty \!\! d\tau {[1+\tau (\mu\xi + \eta (1-\mu))]^{2\omega-5}\over
(1+\tau)^{\omega}}\nonumber \\
&&\times
{[(x-z)_+ +\tau\eta (1-\mu) (x-y)_+][(y-z)_+ + \tau \mu\xi
(y-x)_+]^2\over[-\mu\xi(x-z)^2 -\eta (1-\mu) (y-z)^2 - \tau\xi\eta\mu(1-\mu)
(x-y)^2 +i\e]^{2\omega-1}}\ .
\end{eqnarray}
Since  ${\cal V}_1$ has an explicit zero at $\omega=1$, if we show that the
integral in (\ref{gamma1}) is convergent when evaluated at $\omega=1$, we
have proved that the three point Green function vanishes at $D=2$. 
Integral (\ref{gamma1}) is discussed in the Appendix.

\section{Conclusions}

A peculiar feature of the light-cone gauge formulation  of
Yang-Mills theories  is that they can be consistently defined
in two dimensions: contrary to the covariant Feynman  gauge, 
the light-cone gauge
propagator with ML prescription for the spurious pole is a 
tempered distribution
at $D=2$. In particular, the large $T$ behaviour of the Wilson loop 
can be evaluated without the need of introducing any regulator;
the finite result has been presented in ref.\cite{Bas7}. This result, 
however, cannot be compared with the result one would obtain in Feynman
gauge, as in the latter case, the free propagator is not a tempered 
distribution at $D=2$. In Feynman gauge the best one can do is to
evaluate the Wilson loop in $D$ dimensions, and to take eventually 
the limit $D\to 2$. 

In so doing one obtains again a finite result \cite{bello} that, however, 
is {\it different}
from the one of ref.\cite{Bas7}. In LCG the diagram with a self-energy 
correction in the propagator, which only exists in $D>2$, makes the 
difference. It is precisely the contribution we have evaluated in this paper.
It provides us with the missing term to get agreement
between refs. \cite {Bas7} and \cite{bello}, {\it i.e.} to recover
gauge invariance. Such a phenomenon was not unexpected in the light of
ref.\cite{Bas1}. Perturbative Yang-Mills theory in LCG looks indeed 
discontinuous in the limit $D\to 2$; actually, starting from a
vanishing coupling at $D=2$, it exhibits a kind of 
``instability'' with respect to a change of dimensions. 

On one hand our result clarifies the nature of the discontinuity of
Yang--Mills theories in two dimensions, on the other it rises new
interesting questions for future investigations. 

While in any dimension $D>2$ 
$perturbative$ Wilson loop calculations are in agreement with 
Abelian-like time exponentiation, as all $C_A$ dependent terms 
turn out to be depressed in the large-$T$ limit, at $D=2$
neither the result in ref. \cite{Bas7} nor the one in \cite{bello} share
this property, as they
both exhibit an explicit $C_A$-dependence in the coefficient of the
leading term when $T \to \infty$. At $D=2$
exponentiation in terms of $C_F$
occurs perturbatively only in light-front formulation (ref. \cite{thooft});
in equal-time quantization, exponentiation requires full resummation
of genuine non-perturbative contributions (instantons) \cite{basgri}.

The difference between the formulations above (and their related vacua)
as well as the reason why this phenomenon seems to be crucial only at $D=2$
are under active investigation.

\appendix
\section*{}

In this appendix we show that the three point Green function tends to
zero when $D\to 2$. As explained in the main text, it is sufficient to prove
that the integral in
 eq. (\ref{gamma1}),
with  the constant containing the  simple zero $(\omega -1)$ factorized out, is
convergent when evaluated at $\omega =1$. Such an integral,
 after the change of variables $\alpha = \mu\xi$, $\beta=\eta (1-\mu)$
and after explicit integration over  $d\mu$, reads
\begin{eqnarray}
\label{integral}
&&I=\int_0^1  d\alpha d\beta \theta (1-\alpha -\beta)\left\{{1-\alpha-\beta\over
1-\alpha}+\beta\log{(1-\beta)(1-\alpha)\over \alpha\beta}\right\}\int_0^\infty
{d\tau\over(1+\tau)}\nonumber\\
&\times & {[(x-z)_+ +\beta\tau (x-y)_+]\over [1+\tau (\alpha+\beta)]^3}{[(y-z)_+
+\alpha\tau (y-x)_+]^2\over [-\alpha (x-z)^2 -\beta (y-z)^2 - \alpha\beta \tau
(x-y)^2 + i\epsilon]}\ ,
\end{eqnarray}
$\theta$ being the Heavyside function. The most delicate region of this
integral
is $\alpha\sim\beta\sim 0$, so that in order to check
convergence of eq. (\ref{integral}) we can restrict ourselves  to the case when
the curly bracket is replaced by one. After this replacement, we set
$\alpha=\rho\sigma$ and $\beta = \rho (1-\sigma)$.
In the expression obtained after this change of variables,  we rescale
$\gamma=\rho\tau$ at fixed $\tau$.
The integral over the $\tau$ variable can be
factorized providing a factor $\log(1 + 1/\gamma)$.
Finally, renaming $\rho=1/\gamma$, eq. (\ref{integral}) with the curly bracket
replaced by one can be equivalently written as
 \begin{equation} \label{integral2}
{\cal I}=-\int_0^1\!\! d\sigma\int_0^\infty \!\!
{d\rho\over\rho}{\log(1+\rho)\over(1+\rho)^3}{[\rho(x-z) +
(1-\sigma)(x-y)]_+[\rho (y-z) +\sigma (y-x)]_+^2\over [\rho\sigma (x-z)^2 +\rho
(1-\sigma )(y-z)^2 +\sigma(1-\sigma) (x-y)^2-i\epsilon]}\ .
\end{equation}
Dividing the $\rho$ integration domain as $[0,1]\cup [1,\infty)$, we split
${\cal I}$ as ${\cal I}_1 + {\cal I}_2$. In ${\cal I}_1$, $\rho\in[0,1]$ and
therefore we can use the following majorations: $\log(1+\rho)<\rho$ and
$(1+\rho)^{-3}<1$. Thus, integration   in $d\rho$ is straightforward, providing
us with the estimate
\begin{eqnarray}
&&{\cal I}_1 \simeq -\int_0^1 d\sigma {(x-z)_+ (y-z)^2_+\over \sigma (x-z)^2 +
(1-\sigma) (y-z)^2 -i\epsilon}\nonumber\\
&& \left[ {1\over 3} + {1\over 2} (A-C) +
(B-C)^2 + B + 2AB - AC + (A-C)(B-C)^2 \log\left( {1+C\over C}\right)\right]\ ,
\end{eqnarray}
where $A$, $B$ and $C$ are defined as
\begin{eqnarray}
A &=& (1-\sigma)(x-y)_+/(x-z)_+ \ ,\nonumber\\
B &=& \sigma (x-y)_+/(z-y)_+ \ ,\nonumber\\
C &=& \sigma (1-\sigma)(x-y)^2/[\sigma (x-z)^2 + (1-\sigma)(y-z)^2
-i\epsilon]\ .
\end{eqnarray}
In this form, it is manifest that integration over $\sigma $ is convergent. The
explicit result goes beyond the purpose of the paper, but it can be easily
evaluated providing combinations of rational functions, logarithms and
dilogarithms.

In  ${\cal I}_2$, the $\rho $ integration domain is $[1,\infty)$ and therefore
we can use $(1+\rho)^{-3} < \rho^{-3}$. Thus, the $\rho$ dependent part of the
integrand can be approximated by
\begin{eqnarray}
\label{alg}
&&{(\rho + A)(\rho+B)^2\over (\rho + C)}
{\log(1+\rho)\over \rho^4}=\nonumber\\
&&{\log(1+\rho)\over \rho (\rho+C)}+ {A (\rho +B)^2 + \rho (B^2 + 2 \rho B)\over
(\rho +C)\rho^3} {\log(1+\rho)\over \rho}
\end{eqnarray}
To check convergence, in the second term of the r.h.s. we can replace
$\log(1+\rho)/\rho$ by $1$. Then, integration over $\rho$ becomes
straightforward  and the second term in eq. (\ref{alg}) provides integrals over
$d\sigma$ of the same kind of those in ${\cal I}_1$, where convergence can be
easily checked. The first term in the r.h.s. of eq. (\ref{alg}) is
more delicate. Here  the majoration $\log(1+\rho)<\rho$ is too strong as it
would spoil convergence in  the $\rho$ integration. An explicit integration
over $\rho$ of this term gives
\begin{eqnarray}
\label{iuno}
&&{\cal I}_2^{first}\simeq \int_0^1 d\sigma{(x-z)_+(y-z)_+^2\over \sigma (x-z)^2 +
(1-\sigma) (y-z)^2 -i\epsilon}\times\nonumber\\
&&{1\over C}\left[ {\rm Li} \left({ C\over C-1}\right) + {\rm Li} \left(- C
\right) - \log 2 \log \left( {1+C\over 1-C} \right) - {\rm Li}
\left({2C\over C-1}\right)\right]\ , \end{eqnarray}
${\rm Li} (z)$ being the dilogarithm function. Although cumbersome, integration
over $\sigma $ is finite.


\begin{references}

\bibitem{wit}{E. Witten, Commun. Math. Phys. \underbar{141}, 153 (1991) and J.
Geom. Phys. \underbar{9}, 303 (1992).}

\bibitem{doug}{M.R. Douglas and V.A. Kazakov, Phys. Lett. \underbar{B319},
219 (1993);\\
D.V. Boulatov, Mod. Phys. Lett. \underbar{A9}, 365 (1994);\\
D.J. Gross and A. Matytsin, Nucl. Phys. \underbar{B429}, 50 (1994); {\it ibid}
\underbar{B437}, 541 (1995).}

\bibitem{basgri}{A. Bassetto and L. Griguolo, Phys.
Lett. \underbar{B443} 325 (1998).}


\bibitem{poly72}{A. M. Polyakov, Phys. Lett. \underbar{82B}, 247 (1979);\\
J. B. Kogut and L. Susskind, Phys. Rev. \underbar{D11}, 395 (1975).}

\bibitem{fish}{W. Fishler, Nucl. Phys. \underbar{B129}, 157 (1977).}

\bibitem{wils74}{K. Wilson, Phys. Rev. \underbar{D10}, 2445 (1974);\\
L.S. Brown and W.I. Weisberger, {\it ibid} \underbar{20}, 3239 (1979).}

\bibitem{tellu}{A. Bassetto and G. Nardelli, Int. J. Mod. Phys. \underbar{A12},
1075 (1997) and Erratum-ibid. \underbar{A12}, 2947 (1997).}

\bibitem{thooft}{ G. 't Hooft, Nucl. Phys. \underbar{B75}, 461
(1974).}

\bibitem{call76}{ C.G. Callan, N. Coote and D.J. Gross, Phys.Rev.
\underbar{D 13}, 1649 (1976).}

\bibitem{Bas4}{ A. Bassetto, G. Nardelli and R. Soldati, {\it Yang--Mills
theories in algebraic non covariant gauges} (World Scientific,
Singapore, 1991).}

\bibitem{Wu}{ T.T. Wu, Phys. Lett. \underbar{71B}, 142
(1977).}

\bibitem{ML}{S. Mandelstam, Nucl. Phys. \underbar{B213}, 149 (1983);\\
G. Leibbrandt, Phys. Rev. \underbar{D29}, 1699 (1984).}

\bibitem{Bas5}{A. Bassetto, M. Dalbosco, I. Lazzizzera and R. Soldati,
Phys. Rev. \underbar{D31}, 2012 (1985).}

\bibitem{Bas3}{A. Bassetto, M. Dalbosco and R. Soldati, Phys. Rev.
\underbar{D36}, 3138 (1987).}



\bibitem{Korc}{ I.A. Korchemskaya and G.P. Korchemsky,
Phys. Lett. \underbar{B 287}, 169 (1992).}

\bibitem{Bas2}{A. Bassetto, I.A. Korchemskaya, G.P. Korchemsky and G.
Nardelli, Nucl. Phys. \underbar{B 408}, 62 (1993);\\
A. Andrasi and J. C. Taylor, Nucl. Phys. \underbar{B357},
341 (1992); [Erratum-ibid. \underbar{B414}, 856 (1994)]. }

\bibitem{Bas1}{A. Bassetto, F. De Biasio and L. Griguolo,
Phys. Rev. Lett. \underbar{72}, 3141 (1994).}


\bibitem{bello}{A. Bassetto, R. Begliuomini and G. Nardelli,  Nucl. Phys.
 \underbar{B534}, 491 (1998).}

\bibitem{ALTE}{S. Caracciolo, G. Curci and P. Menotti, Phys. Lett.
\underbar{113B}, 311 (1982);\\
J.P. Leroy, J. Micheli and G.C. Rossi, Nucl. Phys. \underbar{B232},
511 (1984).}

\bibitem{stau}{M. Staudacher and W. Krauth, Phys. Rev. \underbar{D57},
2456 (1998).}

\bibitem{Bas7}{A. Bassetto, D. Colferai and G. Nardelli, Nucl. Phys.
\underbar{B501}, 227 (1997) and Erratum-ibid  \underbar{507}, 746
(1997).}

\bibitem{tay}{J. Frenkel and J. C.Taylor, Nucl. Phys. \underbar{B246},
 231 (1984).}

\end{references}
\end{document}